\RequirePackage[orthodox,l2tabu]{nag}
\documentclass{llncs}

\usepackage[english]{babel}
\usepackage[utf8]{inputenc}
\usepackage[T1]{fontenc}
\usepackage[rm={oldstyle=false,proportional=true},sf={oldstyle=false,proportional=true},tt={oldstyle=false,proportional=true,variable=true},qt=false]{cfr-lm}
\usepackage{zi4}
\usepackage{amsmath,graphicx,tikz-cd}
\usepackage{svg,graphicx,xcolor}
\usepackage{csquotes}
\usepackage{cite}
\usepackage{url}
\usepackage{listings}
\usepackage{mdframed}
\usepackage{ifthen}
\usepackage[export]{adjustbox}[2011/08/13]
\usepackage{authblk}

\usepackage{tikz}
\usetikzlibrary{shapes}

\usepackage{varioref}
\usepackage[pdfpagelabels=true]{hyperref}
\usepackage{cleveref}

\hypersetup{
	linktoc = page,
	pdfpagemode = UseNone,
	colorlinks,
	linkcolor={red!50!black},
	citecolor={blue!50!black},
	urlcolor={blue!80!black}
}

\lstdefinelanguage
   [x64]{Assembler}     
   [x86masm]{Assembler} 
   {morekeywords={CDQE,CQO,CMPSQ,CMPXCHG16B,JRCXZ,LODSQ,MOVSXD, %
                  POPFQ,PUSHFQ,SCASQ,STOSQ,IRETQ,RDTSCP,SWAPGS, %
                  rax,rdx,rcx,rbx,rsi,rdi,rsp,rbp, %
                  r8,r8d,r8w,r8b,r9,r9d,r9w,r9b,cmove}}

\newcommand{\CFG}[1]{\operatorname{CFG}(#1)}

\begin{document}

\title{Generating Functionally Equivalent Programs Having Non-Isomorphic Control-Flow Graphs}
\author{Rémi Géraud$^1$ \and Mirko Koscina$^{12}$ \and Paul Lenczner$^1$ \and\\  David Naccache$^1$ \and David Saulpic$^1$}
\institute{$^1$\'Ecole normale sup\'erieure\\
45 rue d'Ulm, \textsc{f}-75230 Paris \textsc{cedex 05}, France\\
\email{\url{given\_name.family\_name@ens.fr}} \\
$^2$Almerys\\
46 Rue du Ressort, 63967 Clermont-Ferrand \textsc{cedex 9}, France\\
\email{\url{mirko.koscina@almerys.com}}
}

\maketitle

\begin{abstract}
One of the big challenges in program obfuscation consists in modifying not only the program's straight-line code (SLC) but also the program's \emph{control flow graph} (CFG). Indeed, if only SLC is modified, the program's CFG can be extracted and analyzed. Usually, the CFG leaks a considerable amount of information on the program's structure.

In this work we propose a method allowing to re-write a code $P$ into a functionally equivalent code $P'$ such that $\CFG{P}$ and $\CFG{P'}$ are radically different.
\end{abstract}

\section{Introduction}

In the white-box security model, adversaries have access to a program's internals --- assembly code, memory, etc. This model captures real-world attacks against low-end devices, as well as software disassembly and dynamic analysis. Such attacks may allow the adversary to extract secrets from the implementation, either in the form of tokens (passwords, etc.), intellectual property (algorithms, etc.), or may help uncover design flaws that may later be exploited. Reverse-engineering may also help the adversary recognize some trait that the program shares with other programs, e.g. in the case of malware analysis or intellectual property infringement. The general aim of obfuscation is to prevent reverse-engineering, by defeating automated methods and stave off human efforts to make sense of the code. Applications of RE-evasion techniques are many, and constitute for instance an essential building block of digital rights management (DRM) systems.

Historically, program identification focused on finding known code chunks called \emph{signatures} in the binary.
While this technique is still widely in use amongst intrusion and virus detection systems, such an approach requires both extensive, and up-to-date, databases (to account for the ever-growing corpus of threats) and a very efficient binary comparison method. At the same time, widely used packagers with self-modifying code capacity, now standard amongst virus designers, made the traditional signature-based  approach less and less effective. 

Indeed, an increasing number of malicious programs re-write their executable code so as not to feature any recognizable code of significant length. In practice, it is not even necessary to resort to very complex re-writing mechanisms: the malicious code can simply add (or remove) useless instructions or instruction sequences (such as \texttt{nop}, and reversible register operations, e.g. \texttt{inc}/\texttt{dec}) to thwart a trivial comparison. While such variations can be accounted for, they require significantly more effort from the analyst, especially when scanning a large number of files.

An alternative, and certainly complementary approach to malware detection and analysis consists in running the program under certain controlled environment, or \emph{sandbox}, in which every operation can be monitored and does not impact the ``real'' underlying system. Sandboxes typically implement a form of virtualized environment, and monitor access to resources, secrets, and peripherals to detect abnormal behavior. Naturally, the term \enquote{abnormal} is application-dependent, hence this approach assumes that characteristic behavioral features are known and are sufficiently distinguishable from those of uninfected software. Furthermore, running such a controlled environment is resource- and time- demanding. This limits the interest of sandboxing as a program identification tool. 

Between these two approaches, recent research focused on methods for comparing programs using control-flow graph isomorphism \cite{SSTIC:DulRol2005,DIMVA:Flake2004,RAID:KKMRV2005}. The rationale is that the program's flow graph (CFG) wouldn't be altered significantly by the adjunction or removal of useless ``decoy'' operations, the kind of which thwarts direct comparison. CFG comparison techniques are also unaffected by straight-line code obfuscation techniques, e.g. when each function's code is completely rewritten. CFGs can be extracted statically to a large extent, and therefore constitute an attractive and resource-frugal alternative to full-blown virtualisation.

\paragraph{Defeating CFG analysis.} In the malware-writing community, a typical anti-reverse engineering technique is the \emph{trampoline}: instead of using typical control flow instructions such as \texttt{jmp} or \texttt{call}, the program makes heavy use of exception handling, that preempts the instruction pointer and runs the exception handler, which re-dispatches control flow to another program part (see e.g. \cite{davi2015code}). After execution, each program part raises an exception, and falls back to an exception handler (hence the name, trampoline). There can be several trampolines, which may be created and moved at runtime, and code boundaries need not be rigid. This prevents disassemblers from reliably cross-referencing information, and makes it difficult to perform dynamic analysis as well, because it is typically impossible to run such code within a debugger. 

However, because there is no classical call hierarchy, trampolines have to emulate the stack, and an analyst that recognizes the mechanism can easily reconstruct the control flow graph by following this pseudo-stack. Therefore, while the use of trampolines slows down analysis, it is by no means an efficient method anymore against trained reverse engineers, and the additional effort put into designing such code is not worth the marginal gain.

Recent work tried to automate the process, which strives to achieve a ``flat'' control flow graph, i.e. a graph with either a single central trampoline that dispatches execution, or a program that is fully unrolled and appears as a long straight-line code segment without internal structure \cite{TR:WHKD2000,ICIS:CGJZ2001,CCS:LinDeb2003,misc:PopDebAnd2007,AUSCREN:LasKis2009,DRM:CapPre2010,IWIH:SchKat2011}.
However not only are such techniques not always applicable, but more importantly they tend to produce code that, while ``flat'', has salient signatures.

\paragraph{Our contribution.}
This paper addresses the question of rewriting a program in a way that hides its original control flow graph from static analysis (and, to a certain extent, from dynamic analysis as well), while preserving functionality. Straight-line code (SLC) obfuscation techniques can be used on top of our construction to destroy remaining signatures. Indeed SLC obfuscators have already been described in the literature and shown to effectively defeat classic code analysis techniques \cite{CSUR:SKKMW2016}. The rewriting is randomized, and produces different outputs every time. Unlike the trampoline construction, whose heavy use of exception handling is easily recognizable, and from there, traceable, our construction only uses common instructions and relies on a specific routing mechanism along execution --- which is much harder to detect. 

More formally, given a program $P$, we show how to obtain a functionally equivalent program $P'$, such that the CFG of $P'$ is essentially a random graph. This transformation is automatic, and we show how to implement a CFG-transcompiler for the x86-64 architecture, which is widely used and furthermore makes our implementation easier.
   
\section{Control Flow Graph Transcompilation}
\subsection{Prerequisites}
The control flow graph of a program is a graphical representation, based on nodes and edges, of the paths that might be traversed by the program during its execution.
\begin{definition}[Control Flow Graph]
The (full) \emph{control flow graph} of a program $P$ is the graph whose nodes are the program's instructions and the edges are control flow transitions. The \emph{restricted control flow graph} of $P$ has for nodes \emph{straight-line blocks}, i.e. a maximal sequence of code without departure or arrival of static jumps, and there is an edge from node $x$ to node $y$ (and we write $x \to y$) if either of the following conditions hold:
\begin{itemize}
    \item The code of node $y$ is located immediately after the node $x$, and both are separated by a conditional jump.
    \item The last instruction of the node $x$ is either a conditional or a static jump, which is a call to the physical address of the beginning of the node $y$.
\end{itemize}
\end{definition}
In the following, unless specified otherwise, we always refer to the \emph{restricted} control flow graph.
This construction does not include information about dynamic jumps: In practice it is challenging to statically and reliably resolve dynamic jumps. The \texttt{ret} instruction, which we cannot ignore since it is often used to implement function calls, will be dealt with in a special way. 

However, other dynamic and indirect control flow modifications (e.g. by direct alteration of the instruction pointer, or non-standard exception handling) are not considered in this work. On the one hand this is a limitation that may prevent some programs from undergoing the transformation that we propose. On the other hand, this may constitute an interesting countermeasure against code-reuse and hijack attacks that leverage such possibilities.

Let $P$ be the program to be obfuscated. We denote by $G = (V, E)$ the CFG of $P$, where $V$ and $E$ correspond respectively to the nodes and edges of $G$. Let $G' = (V', E')$ be a given \enquote{final} target CFG. 

\begin{example}\label{example:cfg}
Consider the following program, implementing a simple double-and-add algorithm:
\begin{lstlisting}[language={[x64]Assembler}]
dbl_add(int, int):                  ; Compute ab from integer arguments a and b 
        test    esi, esi
        mov     eax, 0              ; tmp = 0
        jle     .end                ; if b == 0, return tmp
.loop:
        lea     edx, [rax+rax]      ; tmp2 = 2 tmp
        add     eax, edi            ; tmp = tmp + a
        test    sil, 1
        cmovne  eax, edx            ; if b even set tmp = tmp2
        sar     esi                 ; shift b to the right
        jne     .loop               ; loop if b > 0
        rep ret
.end:
        rep ret
\end{lstlisting}
The CFG associated to this program is represented in \Cref{fig:cfg1}, where
the instructions' arguments have been removed for clarity. The associated restricted CFG is represented in \Cref{fig:cfg2}.
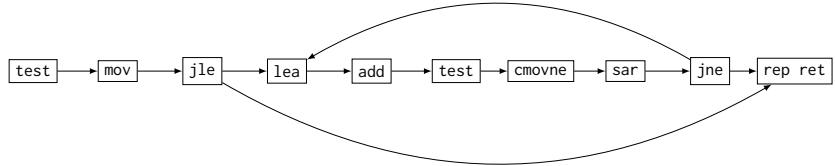
\begin{figure}[!ht]
\centering \scalebox{0.75}{
\begin{tikzpicture} 
\node[draw, rectangle] (1) at (0, 0) {\texttt{test}};
\node[draw, rectangle] (2) at (1.5, 0) {\texttt{mov}};
\node[draw, rectangle] (3) at (3, 0) {\texttt{jle}};
\node[draw, rectangle] (4) at (4.5, 0) {\texttt{lea}};
\node[draw, rectangle] (5) at (6, 0) {\texttt{add}};
\node[draw, rectangle] (6) at (7.5, 0) {\texttt{test}};
\node[draw, rectangle] (7) at (9, 0) {\texttt{cmovne}};
\node[draw, rectangle] (8) at (10.5, 0) {\texttt{sar}};
\node[draw, rectangle] (9) at (12, 0) {\texttt{jne}};
\node[draw, rectangle] (10) at (13.5, 0) {\texttt{rep ret}};
\draw[->, >=latex] (1) edge (2);
\draw[->, >=latex] (2) edge (3);
\draw[->, >=latex] (3) edge (4);
\draw[->, >=latex] (4) edge (5);
\draw[->, >=latex] (5) edge (6);
\draw[->, >=latex] (6) edge (7);
\draw[->, >=latex] (7) edge (8);
\draw[->, >=latex] (8) edge (9);
\draw[->, >=latex] (9) edge (10);
\draw[->, >=latex] (9) edge[bend right] (4);
\draw[->, >=latex] (3) edge[bend right] (10);
\end{tikzpicture}}
\caption{Full CFG of the program of \Cref{example:cfg}.}\label{fig:cfg1}
\end{figure}

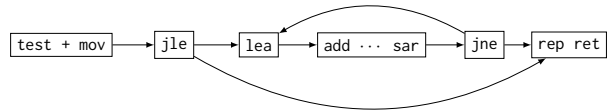
\begin{figure}[!ht]
\centering \scalebox{0.75}{
\begin{tikzpicture} 
\node[draw, rectangle] (1) at (1, 0) {\texttt{test + mov}};
\node[draw, rectangle] (3) at (3, 0) {\texttt{jle}};
\node[draw, rectangle] (4) at (4.5, 0) {\texttt{lea}};
\node[draw, rectangle] (5) at (6.5, 0) {\texttt{add $\cdots$ sar}};
\node[draw, rectangle] (9) at (8.5, 0) {\texttt{jne}};
\node[draw, rectangle] (10) at (10, 0) {\texttt{rep ret}};
\draw[->, >=latex] (1) edge (3);
\draw[->, >=latex] (3) edge (4);
\draw[->, >=latex] (4) edge (5);
\draw[->, >=latex] (5) edge (9);
\draw[->, >=latex] (9) edge (10);
\draw[->, >=latex] (9) edge[bend right] (4);
\draw[->, >=latex] (3) edge[bend right] (10);
\end{tikzpicture}}
\caption{Restricted CFG of the program of \Cref{example:cfg}.}\label{fig:cfg2}
\end{figure}
\end{example}

\subsection{Overview of our Approach}
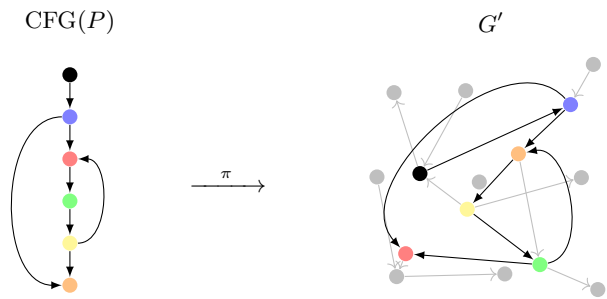
\begin{figure}[!ht]
\centering 
\begin{tikzpicture}[node/.style={circle,fill=black,inner sep=0pt,minimum size=2mm},bcknode/.style={circle,fill=gray!50,inner sep=0pt,minimum size=2mm},bluenode/.style={circle,fill=blue!50,inner sep=0pt,minimum size=2mm},rednode/.style={circle,fill=red!50,inner sep=0pt,minimum size=2mm},greennode/.style={circle,fill=green!50,inner sep=0pt,minimum size=2mm},yellownode/.style={circle,fill=yellow!50,inner sep=0pt,minimum size=2mm},orangenode/.style={circle,fill=orange!50,inner sep=0pt,minimum size=2mm},scale=1.4]
\node (1) at (0,0) [node] {};
\node (2) at (0,-0.4) [bluenode] {};
\node (3) at (0,-0.8) [rednode] {};
\node (4) at (0,-1.2) [greennode] {};
\node (5) at (0,-1.6) [yellownode] {};
\node (6) at (0,-2) [orangenode] {};
\node at (0, 0.5) {$\CFG{P}$};
\node at (4, 0.5) {$G'$};
\node at (1.5, -1) {$\xrightarrow{\quad\pi\quad}$};

\foreach \i in {0,...,2}
    \foreach \j in {0,...,2}
          \node [bcknode] (n-\i\j) at (3+2*.5*\i+0.25*rand,-2*0.5*\j+0.25*rand) {};

\node [node] (m1) at (2.75+.5*1.25+0.125*rand,-0.5*2.5+0.075*rand+0.25) {};
\node [bluenode] (m2) at (2.75+.5*4+0.075*rand,-0.5*1+0.075*rand+0.25) {};
\node [orangenode] (m3) at (2.75+.5*3+0.075*rand,-0.5*2+0.075*rand+0.25) {};
\node [yellownode] (m4) at (2.75+.5*2+0.075*rand,-0.5*3+0.075*rand+0.25) {};
\node [greennode] (m5) at (2.75+.5*3.5+0.075*rand,-0.5*4+0.075*rand+0.25) {};
\node [rednode] (m6) at (2.75+.5*1+0.075*rand,-0.5*4+0.075*rand+0.25) {};

\draw[->,gray!50] (n-01) -- (n-02);       
\draw[->,gray!50] (m1) -- (n-00);        
\draw[->,gray!50] (n-10) -- (m1);  
\draw[->,gray!50] (n-20) -- (m2);
\draw[->,gray!50] (m5) -- (n-22);
\draw[->,gray!50] (m3) -- (m5);
\draw[->,gray!50] (m4) -- (m1);
\draw[->,gray!50] (m4) -- (n-21);
\draw[->,gray!50] (m6) -- (n-02);
\draw[->,gray!50] (n-02) -- (n-12);
        
\draw[->, >=latex] (1) edge (2);
\draw[->, >=latex] (2) edge (3);
\draw[->, >=latex] (3) edge (4);
\draw[->, >=latex] (4) edge (5);
\draw[->, >=latex] (5) edge (6);
\draw[->, >=latex] (5) edge[bend right=90] (3);
\draw[->, >=latex] (2) edge[bend right=90] (6);

\draw[->, >=latex] (m1) -- (m2);
\draw[->, >=latex] (m2) -- (m3);
\draw[->, >=latex] (m3) -- (m4);
\draw[->, >=latex] (m4) -- (m5);
\draw[->, >=latex] (m5) -- (m6);
\draw[->, >=latex] (m5) to[bend right=90] (m3);
\draw[->, >=latex] (m2) to[bend right=90] (m6);
\end{tikzpicture}
\caption{Illustration of Step 1: Relabeling. The original nodes and edges from $\CFG{P}$ are assigned different colors, other nodes are in gray.}\label{fig:step1}
\end{figure}

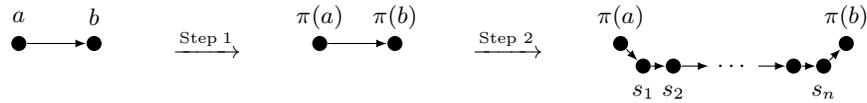
\begin{figure}[!ht]
\centering 
\begin{tikzpicture}[node/.style={circle,fill=black,inner sep=0pt,minimum size=2mm}]
\node[node] (1) at (0,0) {};\node[node] (2) at (1,0) {};
\node at (0,0.35) {$a$};\node at (1,0.35) {$b$};
\node at (2.5,0) {$\xrightarrow{\text{Step 1}}$};
\node[node] (3) at (4,0) {};\node[node] (4) at (5,0) {};
\node at (4,0.35) {$\pi(a)$};\node at (5,0.35) {$\pi(b)$};
\node at (6.5,0) {$\xrightarrow{\text{Step 2}}$};
\node[node] (5) at (8,0) {};\node[node] (6) at (11,0) {};
\node at (8,0.35) {$\pi(a)$};\node at (11,0.35) {$\pi(b)$};
\node[node] (5a) at (8.3,-0.3) {};\node[node] (5b) at (8.7,-0.3) {};
\node[node] (5c) at (10.3,-0.3) {};\node[node] (5d) at (10.7,-0.3) {};
\node (mid) at (9.5,-0.3) {$\cdots$};
\node at (8.3,-0.65) {$s_1$};\node at (8.7,-0.65) {$s_2$};\node at (10.7,-0.65) {$s_{n}$};
\draw[->, >=latex] (1) -- (2);\draw[->, >=latex] (3) -- (4);
\draw[->, >=latex] (5) -- (5a);\draw[->, >=latex] (5a) -- (5b);\draw[->, >=latex] (5b) -- (mid);
\draw[->, >=latex] (mid) -- (5c);\draw[->, >=latex] (5c) -- (5d);\draw[->, >=latex] (5d) -- (6);
\end{tikzpicture}
\caption{Illustration of Step 2: Breaking edges. The original path $\pi(a) \to \pi(b)$ is extended by a path $f((a,b)) = (s_1, \dotsc, s_n)$ of $G'$.}\label{fig:step2}
\end{figure}

\begin{figure}[!ht]
\centering 
\begin{tikzpicture}[node/.style={circle,fill=black,inner sep=0pt,minimum size=2mm}]
\node[node] (1a) at (0,1) {};\node[node] (1b) at (0,-1) {};\node[node] (2) at (1,0) {};
\node[node] (3) at (2,0) {};\node[node] (4) at (3,0) {};\node[node] (5a) at (4,1) {};
\node[node] (5b) at (4,-1) {};
\node[node,fill=none,draw] (1ab) at (6,1) {};\node[node] (1bb) at (6,-1) {};\node[node,fill=none,draw] (2b) at (7,0) {};
\node[node,fill=gray,draw] (3b) at (8,0) {};\node[node,fill=none,draw] (4b) at (9,0) {};\node[node] (5ab) at (10,1) {};
\node[node] (5bb) at (10,-1) {};
\node at (0,-1.35) {$\pi(a)$};\node at (4,1.35) {$\pi(b)$};
\node at (2,-0.35) {$\pi(c)$};\node at (4,-1.35) {$\pi(d)$};
\node at (5, 0) {$\xrightarrow{\text{Step 3}}$};
\node at (6,-1.35) {$\pi(a)$};\node at (10,1.35) {$\pi(b)$};
\node at (8,-0.35) {$\pi(c)$};\node at (10,-1.35) {$\pi(d)$};
\draw[->, >=latex] (1a) -- (2);\draw[->, >=latex] (1b) -- (2);
\draw[->, >=latex] (2) -- (3);\draw[->, >=latex] (3) -- (4);
\draw[->, >=latex] (4) -- (5a);\draw[->, >=latex] (4) -- (5b);
\draw[->, >=latex] (1ab) -- (2b);\draw[->, >=latex] (1bb) -- (2b);
\draw[->, >=latex] (2b) -- (3b);\draw[->, >=latex] (3b) -- (4b);
\draw[->, >=latex] (4b) -- (5ab);\draw[->, >=latex] (4b) -- (5bb);
\end{tikzpicture}
\caption{Illustration of Step 3: Identifying active and passive nodes. Here two original sequences $\pi(a) \to \pi(b)$ and $\pi(c)\to\pi(d)$ cause some nodes to be passive (empty circle), active (filled black circle), or active  depending on the execution path taken (grey circle).}\label{fig:step3}
\end{figure}
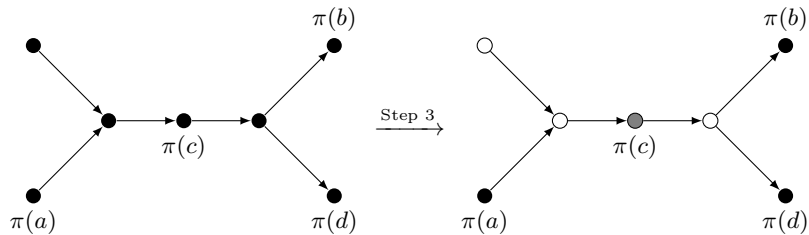

\begin{figure}[!ht]
\centering 
\begin{tikzpicture}[node/.style={circle,fill=black,inner sep=0pt,minimum size=2mm}]
\node[node,fill=none,draw] (1) at (0,1) {};\node[node] (2) at (0,-1) {};
\node[node,fill=none,draw] (3) at (1,0) {};
\node[node,fill=none,draw] (4) at (2,0) {};
\node[node] (5) at (3,0) {};
\node[node] (6) at (4,1) {};\node[node,fill=none,draw] (7) at (4,-1) {};
\node at (0,1.60) {$m=0$};\node at (0,1.35) {$s_1$};
\node at (0,-1.35) {$\pi(a)$};\node at (0,-1.60) {$m=1$};
\node at (1.2,-0.35) {$s_2$};\node at (1.2,-0.60) {$m=0$};
\node at (2,0.35) {$s_3$};\node at (2,0.60) {$m=0$};
\node at (2.8,-0.35) {$\pi(b)$};\node at (2.8,-0.60) {$m=1$};
\node at (4,1.35) {$\pi(c)$};\node at (4,1.63) {$m=1$};
\node at (4,-1.35) {$s_4$};\node at (4,-1.60) {$m=0$};
\draw[->, >=latex] (1) -- (3);
\draw[->, >=latex] (2) -- (3);
\draw[->, >=latex] (3) -- (4);
\draw[->, >=latex] (4) -- (5);
\draw[->, >=latex] (5) -- (6);
\draw[->, >=latex] (5) -- (7);
\end{tikzpicture}
\caption{Illustration of Step 4: The path is taken according the routing variable $m$. If the node is passive ($m=0$), the path to be taken will be the subsequent node. In the case of a active node ($m=1$), the next node will be defined by the current node}
\label{fig:step4}
\end{figure}
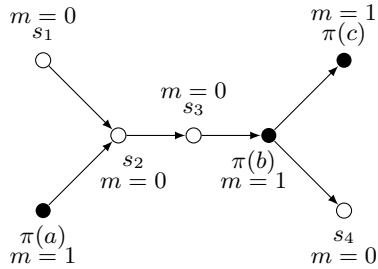

Our goal is to rewrite $P$ into a program $P'$ that achieves the same functionality as $P$, but whose CFG is $G' \not\simeq G = (V, E) = \CFG{P}$. This is achieved in successive steps, illustrated in \Cref{fig:step1,fig:step2,fig:step3,fig:step4}.

\paragraph{Step 1: Relabeling.} We start from a morphism $\pi$ between the two graphs, i.e. a function that is injective on nodes and preserves edges. If we fail to find enough nodes or edges to perform this operation, which happens with very low probability when the target graph is large enough, we simply start over with a new random graph $G'$. The process is illustrated in \Cref{fig:step1}.
\paragraph{Step 2: Breaking Edges.}
Then, additional nodes will be added by transforming the graph. The idea is to replace simple edges by paths in $G' = (V', E')$, i.e. for each edge $(a, b) \in E$, corresponding to an edge $(\pi(a), \pi(b))\in E'$, we replace $(\pi(a), \pi(b))$ by a path $\left(\pi(a), f((a,b)), \pi(b)\right)$, where $f$ is a prescribed function. Such a function $f: E \to \operatorname{List}(V')$ must return paths already present in $G'$, i.e. assuming that $f((a,b)) = (s_1,\dotsc, s_n)$, 
\begin{itemize}
\item $(\pi(a), s_1) \in E'$
\item $(s_n, \pi(b)) \in E'$
\item $\forall i \in  \{1, \ldots, n-1\}, (s_i, s_{i+1}) \in E'$
\end{itemize}
We keep track of which edges were originally present and which edges were added at this step. The process is illustrated in \Cref{fig:step2}.

\paragraph{Step 3: Identify Active and Passive Nodes.}
The previous step introduced \enquote{extra} operations between $a$ and $b$. Since we wish to preserve the original program's functionality, we should make sure that only the original endpoints, $a$ and $b$, are executed, while all the intermediary nodes are without effect when executed. We call $a$ and $b$ the \emph{active} nodes, and the intermediary nodes (i.e. nodes that do not exist in the original CFG) are called \emph{passive}.
\begin{remark}
A node that is neither active nor passive in the control flow graph $G$ can be considered either active or passive in $G'$.
\end{remark}
Depending on the execution path taken, some nodes may be active or passive (e.g. \Cref{fig:step3}). To decide whether a given node is active or passive, the program (more precisely, the node itself) checks at runtime the value of a routing variable (see below).

\paragraph{Step 4: Routing.}
Finally, we transform each node so that the execution of passive nodes is without side effects (a process we call \emph{passivation}), except continuing through the sequence of nodes until an active node is attained. To that end we introduce an additional \enquote{routing} variable that will be updated as the program is executed (e.g. \Cref{fig:step4}).

Nodes consult the routing variable to know whether they are active or not; if not, they simply hand over execution to the next node in sequence (possibly after executing dummy instructions).

\subsection{Contexts}
During program execution, every node in the transformed program undergoes the following procedure:
\begin{enumerate}
\item Determine whether node is active or passive.
\item If active, restore the registers. Otherwise passivate itself.
\item Run the code.
\item Call the next node in the sequence.
\end{enumerate}
To allow this series of operations, we introduce the concept of \emph{contexts}.

A context is a set of variables that save the node's state, in a way that can later be restored. Each traversed node is associated to a context, which is available just during the time that the node is being traversed.



Since passive nodes do not suffer side effects, they cannot in particular find the next node to be called; hence the next node is part of the context. If the node is active, it may ignore this part of the context and branch itself to another destination.

\subsection{Node Passivation}
Node passivation requires us to cancel the instruction(s) being executed, or compensate their effects in some way. We do this by using both the registers and the stack (it is not possible to rewrite registers that are in active mode), leveraging the specificities of the x86-64 architecture.

\paragraph{Register operations.} Any register operation can be dealt with by using contexts, with the exception of the stack registers. 

\paragraph{Stack operations.} Stack operations are harder to compensate: the following instructions have an effect on the stack
\begin{center}
\texttt{PUSH, POP, PUSHA, POPA, PUSHAD, POPAD, PUSHF, POPF, PUSHFD, POPFD}
\end{center}
We control writing and reading in the stack by using a pointer to a \enquote{trash} address, stored as a fixed value. If a passive node attempts to write something in the stack, we redirect the address to the trash, nullifying the instruction's effects. The reading process is handled in the same way. If the node is active, the real address is used.

The context $m$ is used in the following way: after a \texttt{PUSH}, we perform the following operation to the pointer to the top of the stack $p$:
\begin{equation*}
p \gets p + (8\ \&\ m)
\end{equation*}
where
\begin{itemize}
\item $m = 1\cdots1_2$ if the node is passive. In this case the operation will be compensated and will not have any effect due to the top of the stack not changing.
\item $m = 0$ if the node is active. In this case the addition is useless and the \texttt{PUSH} works as intended.
\end{itemize}

\paragraph{\texttt{MOV} instruction.} 
\texttt{mov} instructions from one register to another are already without effect, since register values are restored at the beginning of each active node, and are stored in the environment. However, \texttt{mov} instructions that involve a memory address require additional care, and we use the same technique as for the stack: the address is rewritten to the \enquote{bin} when the node is passive. This is followed by the transformation: 
\begin{equation*}
\text{address} = (\text{address}\ \&\ (\neg m)) | (\text{trash\_address}\ \&\ m)
\end{equation*}
This technique also hides the addresses that are really used during program execution.

\paragraph{Function calls.} 
We will distinguish \emph{library} function calls and calls to \emph{internal} functions, that are defined in the code.

\paragraph{Library calls.} In the case of library function calls, each of them is treated separately by using a specific context per function. Now, considering that it is impossible to handle all the functions at the same time, we propose to call the functions by using parameters that make them ineffective.

\begin{example}
In the following "Hello World" program, where we make ineffective the function printf by loading to EAX the address of an empty sentence (auxiliary parameter) and set the stack pointer to the address of EAX.   
\begin{lstlisting}[language={[x64]Assembler}]
extern _printf
global _main

section .data
param1: db "Hello World",10,0 
paramaux: db "",0               ; declaration of the empty sentence 

section .text
_main:
    push param1
	  lea eax, [paramaux]         ; paramaux address placed in EAX
	  mov [esp], eax              ; pointer to the empty sentence
    call _printf
    add esp,4   
    ret
\end{lstlisting}
\end{example}

\subsection{Jumps and Internal Calls}

\paragraph{Internal calls.} Recall that we distinguish between a call to an address in a \texttt{PUSH} from a static jump. This makes the above transformation effective to handle these instructions. However, the \texttt{RET} instruction corresponds to a dynamic jump and is subtler to handle.

Let $n$ be a node with a \texttt{RET} instruction in $G$, and assume that in $G'$ the corresponding node $\pi(n)$ has two neighbors, $f_1$ and $f_2$. Their addresses are fixed, so that one can place, on the top of the stack, the address of the node that follows $\pi(n)$ (either $f_1$ or $f_2$).

\begin{example}
In the following example, we print on the screen the result returned by \texttt{func1}. In this case, we jump from \texttt{func1} to \texttt{func2} adding the desired address on the top of the stack by using a \texttt{push} operation. As a result, the program jumps to \texttt{func2} instead of jumping back to the address after the call.  
\begin{lstlisting}[language={[x64]Assembler}]
extern _printf
global _main

section .data
num DD 2,3
format: dd "num: %d" , 10, 0

section .text

_main:
	mov eax,0             ; eax = 0
	mov esi, [num]		    ; edi = 2
	mov edi, [num+4]      ; esi = 3
	push esi              ; pass param 3 to .func1
	push edi			        ; pass param 2 to .func1
	push eax              ; pass param 1 to .func1
	call .func1           ; jump to func1
	add esp,12			      ; pop edi, esi and eax from the stack
	
	push eax
	push dword format
    call _printf		    ; print eax in the screen
	add esp,8             ; pop stack 2*4-byte
	
.func1:
	push ebp
	mov ebp,esp 		      ; set stack base pointer
	sub esp, 4            ; creat space for one 4-byte local variable
	push edi			        ; Save the values of the register that the function will use
	push esi
	mov eax,[ebp+8]       ; move param 1 to EAX
	mov edi,[ebp+12]      ; move param 2 to EDI
	mov esi,[ebp+16]      ; move param 3 to ESI
	
	mov [ebp-4],edi       ; var local = 2
	add [ebp-4],esi       ; var local = 5
	mov eax, [ebp-4]      ; EAX = 5 

	pop esi               ; remove esi from the stack
	pop edi				        ; remove edi from the stack
	mov esp,ebp
	pop ebp               ; takedown stack base pointer
	lea	ecx,[.func2]
	push ecx			        ; push func2 address on the top of the stack
	ret                   ; jump func2
	
.func2:
	push ebp
	mov ebp,esp			      ; set stack base pointer
	sub esp, 4            ; creat space for one 4-byte local variable
	push edi              ; Save the values of the register that the function will use
	mov edi,[num]         ; edi = 2
	
	mov [ebp-4],eax       ; var local = 5   
	add [ebp-4], edi      ; var local = 7
	mov eax,[ebp-4]       ; EAX = 7

	pop edi               ; remove edi from the stack
	mov esp,ebp
	pop ebp               ; takedown stack base pointer
	ret
\end{lstlisting}
\end{example}

\subsection{Routing}

Once we have passed through a passive node, without changing the environment, we must be capable to take the next desired branch. As each node is a maximum of two out-degree, all that we need is a boolean variable in the environment that will indicate to which child we must to go. 

In practice, it is enough to maintain a global routing variable $r$. This allows the sequence of branches to follow (left or right) between two consecutive nodes. Hence, we modify $r$ for each active node found and its $i$-th bit gives the direction of the $i$-th branch of the current path. We will denote by $r_i$ the $i$-th bit of $r$.

\begin{remark}
Routing variables have a limited size if we use native types, it is straightforward to extend them but additional arithmetic is needed.
\end{remark}

\paragraph{\texttt{JUMP} instruction.}
First of all, we need to transform a conditional \texttt{jmp} from $P$ into a \texttt{jmp} that goes to the next node as determined by $r_i$. For simplicity, we assume that all conditional jumps test a \enquote{zero flag}, which is set by a comparison just before the jump. For example, we have the node $A$ (with children $B$ and $C$) and the following program:
\begin{lstlisting}[language={[x64]Assembler}]
  cmp (...)     ; comparison
  je B          ; conditional jump to B
  C             ; next node 
\end{lstlisting}
As we know how to move from node $A$ to node $B$ or $C$ in advance, we can save the routes in some constants $A\_to\_B$ and $A\_to\_C$. For doing so we use the following code:
\begin{lstlisting}[language={[x64]Assembler}]
  mov routing_variable, A_to_C   ; set routing variable
  cmp (...)                      ; comparison
  cmove routing_variable, A_to_B ; set routing variable iif comparison succeeds
\end{lstlisting}     
This program then jumps according to the first value of the routing variable.

Note that, for passive nodes, routing variables are set to the (masked) trash address.

\paragraph{\texttt{RET} instruction.}
When a node is passive, we want to have two possible branches as in the case of the jump instruction. To achieve this we also store the constants $A\_to\_B$ and $A\_to\_C$; and we will use the mask $m$ as the context. We will go to node $B$ if $r_i=1$ and to node $C$ if $r_i=0$.

We want to put at the top of the stack the address to which we want to go. Hence, we just add the following line before the \texttt{ret}:
\begin{equation*}
p \gets (p\ \&\ m) | ((r\ \&\ A\_to\_B) | (\neg r\ \&\ A\_to\_C))\ \&\ \neg m
\end{equation*}
The transformation presented above allows us to modify the program's control flow graph. We are capable of transforming an arch into a path, and ensuring that the path's execution is identical to the effect of running the arch in the original graph.

\section{Control Flow Graph Obfuscation}\label{sec:CFGobf}


While the presented construction effectively transforms the program's CFG, the resulting construction has a strong signature, and it is easy to reverse the process to obtain the initial graph. It is indeed enough to run the program and identify nodes that change the routing variable. These nodes are the active ones, and it is possible to reconstruct the original control flow graph.
	
	In this section we propose several ways to obfuscate the transformed program and make this reconstruction harder. First, we will \enquote{force} the execution of the program in order to recover successfully the initial control flow graph, we then hide the nodes' activity, including the operations on the routing variable, which is a signature of an active node.
    
    \subsection{Forcing Execution}
    For now, we know that the routing variable suffices to determine the next active node. We will modify its definition and use it to hide the control flow from static analysis. The routing variable is now maintained as a sequence of bits $(r_1, \dotsc ,r_n)$.
    
    Upon transitioning to node $i$, we apply to the routing bit $r_i$ a random permutation $f_i$ of $\{0,1\}$. 
    \begin{example}
    For example, if one seeks to obtain at the end of the function a bit equal to 1, the following operations can be used:
    \[\begin{array}{ll}
    r \gets 0  & \qquad \text{\color{gray}Null routing variable} \\
    a \gets \operatorname{rand}() & \qquad \text{\color{gray}Introduce randomness} \\
    t \gets 5a \\
    t \gets t + r \times a \\
    r' \gets t/a \bmod 2
    \end{array}\]
   At the end of the code execution we obtain $r' = 1$. If we declare $r=1$ instead, we get $r'=0$. \end{example}
We can easily generate the random flips $f_i$ by using an arithmetic operation and its inverses. As determining the value of a variable is undecidable, running the program is the most natural way to get information about the execution paths taken.

	\subsection{Node Hiding}
    The same way that routing bits are masked, we can hide the value of the bit indicating whether a node is active or passive. However by doing so node $i$ only hides the status of node $i+1$.
	The mask's value can also be changed by choosing a random number between $m$ and $\neg m$, and updating the formula accordingly.
    
	\subsection{Route Hiding}
    Updates of the routing variable are crucial, as they immediately reveal active nodes. 
    To hide the information about the routing changes, we extend each path beyond the active node, and introduce a weak form of \enquote{onion} routing, where the next node is determined at runtime.   
  	The rationale is that determining whether a node is active or inactive will require recovering the full route leading to this particular node.
    
    
	We introduce two additional variables per node, called \emph{path} and \emph{next path}. The \emph{next path} variable is masked (XORed) with a value that depends on the node. Upon execution of an active node, the values of these two new variables are further modified. 
    
    If the next node is $C$, the route from $B$ to $C$ is stored in \emph{next path}, and masked by being XORed with the constants of every intermediary node between $A$ and $B$.
    
    The number of hops is counted. Upon arriving at the final hop $B$ of the path from $A$ to $B$, we swap \emph{next path} and \emph{path}.
    
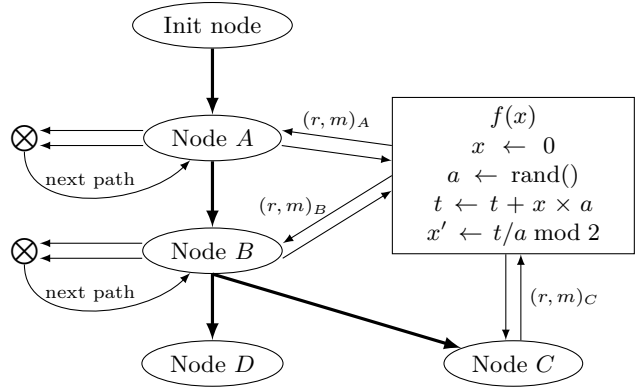
\begin{figure}[!hbt]
\centering
\begin{tikzpicture}[state/.style={ellipse,draw,inner sep=1mm},scale=1]
\node[state] (init) at (0,0) {Init node};
\node[state] (nodeA) at (0,-1.5) {Node $A$};
\node[state] (nodeB) at (0,-3) {Node $B$};
\node[state] (nodeD) at (0,-4.5) {Node $D$};
\node[state] (nodeC) at (4,-4.5) {Node $C$};
\draw[->,>=latex,very thick] (init) -- (nodeA);
\draw[->,>=latex,very thick] (nodeA) -- (nodeB);
\draw[->,>=latex,very thick] (nodeB.south) -- (nodeC);
\draw[->,>=latex,very thick] (nodeB.south) -- (nodeD);
\node[inner sep=0mm] (xor1) at (-2.5,-1.5) {$\bigotimes$};
\node[inner sep=0mm] (xor2) at (-2.5,-3) {$\bigotimes$};
\node[rectangle,draw,align=center,text width=3cm] (block) at (4,-2) {$f(x)$\\$x\gets 0$\\$a \gets \operatorname{rand}()$\\$t \gets t + x \times a$\\$x' \gets t/a \bmod 2$};
\draw[->,>=latex] ([yshift=4mm]block.west) -- node[midway,above] {\scriptsize $(r,m)_A$} ([yshift=1mm]nodeA.east);
\draw[->,>=latex]
([yshift=-1mm]nodeA.east) -- ([yshift=2mm]block.west) ;
\draw[->,>=latex] ([yshift=0mm]block.west) -- ([yshift=1mm]nodeB.east);
\draw[->,>=latex]
([yshift=-1mm]nodeB.east) -- node[midway,above left] {\scriptsize $(r,m)_B$} ([yshift=-2mm]block.west) ;
\draw[->,>=latex]
([xshift=-1mm]block.south) --
([xshift=-1mm]nodeC.north) ;
\draw[->,>=latex]
([xshift=1mm]nodeC.north) --
node[midway,right] {\scriptsize $(r,m)_C$}
([xshift=1mm]block.south)  ;
\draw[->,>=latex]
([yshift=1mm]nodeA.west) --
([yshift=1mm]xor1.east)  ;
\draw[->,>=latex]
([yshift=-1mm]nodeA.west) --
([yshift=-1mm]xor1.east)  ;
\draw[->,>=latex]
([yshift=1mm]nodeB.west) --
([yshift=1mm]xor2.east)  ;
\draw[->,>=latex]
([yshift=-1mm]nodeB.west) --
([yshift=-1mm]xor2.east)  ;
\draw[->,>=latex] (xor1.south) [in=-135,out=-90] to node[midway,above] {\scriptsize next path} (nodeA);
\draw[->,>=latex] (xor2.south) [in=-135,out=-90] to 
node[midway,above] {\scriptsize next path}(nodeB);
\end{tikzpicture}
\caption{Diagram of hiding process for nodes and routes}
\label{fig:digraph}
\end{figure}

    The route hiding process is illustrated in Figure \ref{fig:digraph}. 

\section{Security}

Intuitively, the security of our construction depends on the hardness of identifying active nodes. 
This can be formalized as an adversarial game, whereby a more precise security notion can be given:
\begin{mdframed}\small
\textbf{\underline{CFG-FullRecovery Game}:}
\begin{enumerate}
\item The challenger provides a program CFG $G = (V, E)$ 
\item The adversary chooses a set $N \subseteq V$ 
\end{enumerate}
The adversary wins the game if the nodes in $N$ are the active ones.
\end{mdframed}
To get a grasp on how hard this game is, assume that we choose $N$ at random in $V$, where
there are exactly $|N|$ active nodes:
\begin{equation*}
\Pr\left[ N \text{ is exactly the active nodes} \mid  N \subseteq_R V\right] = \frac{1}{\binom{|V|}{|N|}} = \frac{|N|! (|V| - |N|)!}{|V|!}.
\end{equation*}
If one node out of two is active, and there are more than $42 \times 2$ nodes in $V$, this 
probability is negligible. Thus we may hope, for realistically large programs, to resist
adversaries for which there is no better way to choose $N$ than selecting a random subset of $V$.

However, in practice, adversaries may succeed in recovering smaller portions of the CFG. This corresponds
to the following game:
\begin{mdframed}\small
\textbf{\underline{CFG-OneRecovery Game}:}
\begin{enumerate}
\item The challenger provides a program CFG $G = (V, E)$ 
\item The adversary chooses a node $n \in V$ 
\end{enumerate}
The adversary wins the game if $n$ is active and $n$ is not the first node of $G$ (which is always active).
\end{mdframed}
The success probability if $n$ is chosen at random is
\begin{equation*}
\Pr\left[\text{$n$ is active} \mid n \in_R N \right] = \frac{|N|}{|V|}
\end{equation*}
where again $N$ is the set of (actually) active nodes. In the balanced case, where $2|N| = |V|$ this
probability is exactly one half. When that is the case, and $V$ is large enough, security in this second game implies security in the first game.

As discussed above, static analysis cannot in general determine the variables' values in a given node (by Rice's theorem \cite{rice1953classes}). Given that the difference between active and passive nodes is only semantic, for a general program determining whether a given node is active is undecidable. 

Hence, our obfuscation scheme should be secure against static analysis, for large enough values of $N$ and few enough active nodes.

\subsection{Security Against Dynamic Analysis}
Dynamic analysis is performed by running and monitoring the program. As mentioned previously, the first node is
always active. The second node can be determined as follows: Execution continues until the \emph{next path}
variable is updated. At that point, we know that there is an active node between the current node $B$ and the first node $A$.

The analyst then performs the following operation: For each node $n$ between $A$ and $B$ in the CFG, replace $n$ by another operation, and run the program up to $B$. There are at most $|V|$ nodes to test. A node is active if, when modified, the program's state at $B$ has changed.

As each test is required to continue running the program until $B$, which can take up to $|V|$ steps, it is then possible to determine the next active node in $O(|V|^2)$. By running this procedure iteratively for all nodes, we reconstruct the list of active nodes, i.e. the original CFG, in $O(|V|^3)$ operations.


\section{Implementation}
Given as input a program CFG, we construct a \enquote{target} CFG to which the original program is mapped.

\begin{enumerate}
\item \emph{Graph generation}.
We generate a random graph with $n$ edge and maximum out-dregee two, using a variant of the Tarjan-Eswaran algorithm \cite{eswaran1976augmentation,book:Raghavan2005}. 

\item \emph{Linearisation}. 
This graph is linearised, so that it corresponds to a CFG. For this purpose we use the scheme presented by Leroy for the CompCert compiler \cite{leroy2015compcert}. We then select a random morphism $\pi$ between the initial graph and the new graph that we are creating.

\item \emph{Transformation}. 
We begin the transformation by identifying the active and passive nodes, the edges for paths are then changed by neutralizing (passivation) the instructions using: registers and stacks operations, transforming the jumps and internal call, and defining the route to follow according to the routing variable. Finally, we remove the signature of the new graph hiding the routing variable and node's status (active or passive) by randomizing their values and adding the variables \emph{path} and \emph{next path}. To mask the variable \emph{next path} we XOR it with the node's values.
\end{enumerate}

The source code of this implementation is available from the authors upon request.

\section{Conclusion}
This paper presents a control flow graph trans-compilation algorithm allowing to transforms a program into a new functionally equivalent program. This algorithm uses common instructions such as register and stack operations, and a random routing variable, such that the resulting CFG is entirely different from the original one.
We let as a future work the study of the obfuscation performance regarding the time expended in the transformation process for different code size and if the obfuscation can be improved if we apply on $P$ the same transformation more than once.


\bibliographystyle{plain}
\bibliography{nicfg.bib}

\end{document}